\documentclass[superscriptaddress,12pt]{revtex4-1}

\usepackage[pdftex]{graphicx}
\usepackage{graphicx}
\usepackage{dcolumn}%
\usepackage{bm}%
\usepackage{amsmath} 
\usepackage[caption=false]{subfig}
\usepackage{color}

\usepackage{hyperref}
\usepackage{soul}

\usepackage{color} 
\usepackage{hyperref}
\usepackage{soul}
\usepackage[pdftex]{graphicx}
\usepackage[caption=false]{subfig}
\graphicspath{ {Figures/} }

\begin{document}

\title{Identifying the electron-positron cascade regimes in high-intensity laser-matter interactions}

\author{C. Slade-Lowther}
\affiliation{York Plasma Institute, Department of Physics, University of York, York YO10 5DD, UK}
\author{D. Del Sorbo}
\altaffiliation[Present address: ]{High Energy Density Science Division, SLAC National Accelerator Laboratory, Menlo Park, CA 94025, USA}
\affiliation{York Plasma Institute, Department of Physics, University of York, York YO10 5DD, United Kingdom}
\author{C. P. Ridgers}
\email[Corresponding author: ]{christopher.ridgers@york.ac.uk}
\affiliation{York Plasma Institute, Department of Physics, University of York, York YO10 5DD, UK}

\begin{abstract}
  Strong-field quantum electrodynamics predicts electron-seeded electron-positron pair cascades when the electric field in the rest-frame of the seed electron approaches the Sauter-Schwinger field, i.e. $\eta = E_{RF}/E_S \sim 1$. Electrons in the focus of next generation multi-PW lasers are expected to reach this threshold.  We identify three distinct cascading regimes in the interaction of counter-propagating, circularly-polarised laser pulses with a thin foil by performing a comprehensive scan over the laser intensity (from $10^{23}$ -- $5\times10^{24}$\ Wcm$^{-2}$) and initial foil target density (from $10^{26}$ -- $10^{31}$\ m$^{-3}$). For low densities and intensities the number of pairs grows exponentially. If the intensity and target density are high enough the number density of created pairs reaches the relativistically-corrected critical density, the pair plasma efficiently absorbs the laser energy (through radiation reaction) and the cascade saturates.  If the initial density is too high, such that the initial target is overdense, the cascade is suppressed by the skin effect.  We derive a semi-analytical model which predicts that dense pair plasmas are endemic features of these interactions for intensities above $10^{24}$ Wcm$^{-2}$ provided the target's relativistic skin-depth is longer than the laser wavelength. Further, it shows that pair production is maximised in near-critical-density targets, providing a guide for near-term experiments.
\end{abstract}

%
%
%
%
%

\maketitle

\section{Introduction}

To correctly describe the interaction of strong electromagnetic fields with matter requires strong-field quantum electrodynamics (QED). The well-known ``break down" of the vacuum via pair production is predicted to occur at the critical Sauter-Schwinger field $E_S=m^2 c^3 / e\hbar\approx1.32\times{10}^{18} \, \mathrm{Vm}^{-1}$. Strong-field QED processes can occur in fields far weaker than this critical field. Non-linear Compton scattering of photons in the quantum regime and pair production via the Trident process can occur if the electric field in the rest-frame of the electron (or positron), $E_{RF}$, is equal to the critical field, i.e. the quantum efficiency parameter $\eta = E_{RF}/E_S \sim 1$. We can reach $\eta\sim1$ for laser fields much weaker than $E_S$ as the fields themselves can rapidly accelerate electrons to high Lorentz factor, resulting in a strong Lorentz boost to $E_{RF}$. Pair production by the multi-photon Breit-Wheeler process can occur if the photons emitted during Compton scattering satisfy a similar condition on their quantum efficiency parameter (defined below) $\chi\sim1$. An electromagnetic cascade can ensue if many generations of electrons and positrons can be generated by the fields. Usually this occurs via a two-step process whereby the electrons and positrons produced by the Breit-Wheeler process radiate photons by non-linear Compton scattering which subsequently decay to further pairs and so on.

Upcoming facilities, like several of those comprising the Extreme Light Infrastructure (ELI) \cite{Turcu16,Weber17}, are expected to reach laser intensities of $I > 10^{23} \, \mathrm{Wcm}^{-2}$, and will be capable of accelerating electrons in the plasma generated at the laser focus such that $\eta\sim1$. The possibility of the experimental realisation of this regime, has stimulated investigation of the above QED processes in laser-matter interactions.  In particular, the prediction that strong-field processes might lead to the prolific production of photons and pairs has led to various studies of laser-induced electromagnetic cascades and their requirements \cite{BellKirk08a,Kirk09a,Fedotov10a,Nerush11,Elkina11a,Bashmakov14a,Tang14a,Grismayer17a,Narozhny15a}. These studies suggest that cascades should be possible once laser intensities reach $I \sim 10^{23-24} \, \mathrm{Wcm}^{-2}$, as expected from upcoming facilities. A direct consequence of a pair cascade is the formation of a dense electron-positron pair plasma \cite{Nerush11,Ridgers12a,Ridgers13a,Ji14,Luo15a,Liu16a,Zhu16a,Liu17a}. Pair plasmas generated by cascades are believed to play an important role in extreme astrophysical contexts such as pulsar magnetospheres and active black holes \cite{Goldreich69a,Blandford77a,Timokhin10a}.  Pair plasmas created during a cascade in a laser-plasma interaction are predicted to couple strongly with the field of a laser leading to near-total absorption of the laser pulse \cite{Nerush11,Zhang15a,Grismayer16a}, with consequences for applications of these lasers, for example quenching radiation pressure ion acceleration \cite{Kirk13,DelSorbo18,Duff18}. Hence, the experimental realisation of laser-induced cascades will mark the transition to a regime, as yet only inferred in astrophysical environments, where strong-field QED and plasma effects are coupled \cite{Ridgers12a,Ridgers13a}.  This is in contrast to experiments where non-linear Compton Scattering \cite{Bula96,Sarri14,Cole18,Poder18} and multi-photon Breit-Wheeler pair production \cite{Burke97} have previously been observed in the interaction of electron beams with intense lasers, i.e. not in a plasma environment.

The coupling between plasma and QED effects suggests that pair cascades will be sensitive to the initial target density, plasma effects being far less significant in very low density targets \cite{DelSorbo18}.  In this paper we show that this is indeed the case and that there are three cascade regimes defined by the initial target density (and laser intensity).  This is in contrast to previous work which has usually considered cascades from a small number of seed electrons or very low density plasma \cite{Nerush11,Grismayer17a} or cascades from targets with a narrow range of higher densities, for examples of the latter see Refs. \cite{Ji14,Zhu16a}.  Specifically, we investigate the case of two counter-propagating circularly-polarised laser pulses interacting with a thin foil. Here cascades are seeded by electrons in a plasma target with electron number density $n_0\in[{10}^{26},{10}^{31}] \, \mathrm{m}^{-3}$ for laser pulses each of intensity $I_{24}=(I/({10}^{24} \, \mathrm{Wcm}^{-2}))\in[0.1,5]$. We have simulated cascades using the particle-in-cell (PIC) code \textsc{epoch}, \cite{Arber15a}, which includes the strong-field QED processes described above \cite{Ridgers14a}. 1D \& 2D simulations have been performed in order to show where in $I_{24}$-$n_0$ space cascades and dense pair plasmas will develop, as has previously been done for gamma-ray emission by non-linear Compton scattering only \cite{Brady14a}. In doing so, we outline expectations for future experiments using ultra-high intensity lasers and provide an approximate model for predicting the production of a dense pair plasma over a wide range of possible experimental conditions.

\section{Semi-Analytical Scaling for the Pair Plasma Density}

In order to interpret the cascade simulations presented later, it is useful to derive a simple scaling for the number of pairs produced.  This will also allow us to identify the cascading regimes and their dependence on target density. Similar scalings have been presented previously for cascading from a very low initial electron density \cite{Grismayer17a} and for a cascade in in initially underdense plasma \cite{DelSorbo18,Luo18}.  Here we extend this to include the case of an initially near-critical density or overdense plasma to enable us to explore density space fully. We begin by discussing the electromagnetic fields.

\subsection{Counter-propagating circularly-polarised laser fields}

The electric field of a circularly-polarised plane-wave propagating in the $x$-direction is $\mathbf{E}=E_0f(t)(0,\sin{\phi},\pm \cos{\phi})$, where $\phi=\omega_L t-kx$, $\omega_L$ and $k$ are the laser frequency and wavenumber, respectively, $f(t)$ is a function determining the slowly varying temporal profile of the laser pulse, and $E_0$ is the amplitude of the wave. The $\pm$ sign on the cosine determines the sense of the field rotation. We can further define the laser strength parameter $a_0=eE_0/m\omega_L c$ for later use.\cite{Gibbon05a} For circular-polarisation, $a_0 \approx 600\left(I_{24}\lambda_{\mu m}^2\right)^{1/2}$. A counter-propagating wave is introduced by letting $k\rightarrow -k$. Adding these positive ($+k$) and negative ($-k$) moving components, for same-sense combinations (i.e. the sign on the $z$-component of each beam is $(\pm,\pm)$), the resultant electric field (for ease we set $f(t)=1$) is $\mathbf{E}=2 E_0 \cos{kx} (0,\sin{\omega_L t},\pm \cos{\omega_L t})$ and describes a standing wave in $x$ rotating about the beam-axis, with electric (magnetic) nodes at $kx=n\pi/2$, for odd (even) $n$, i.e. for $x=n\lambda/4$.

If the wavelength is long relative to length scales of processes involved, we can approximate the field by a rotating electric field, given simply by the time-dependent portion of the equation for $\mathbf{E}$ above. In this case, electron motion and pair production can be treated as in \cite{BellKirk08a} to fairly good approximation.

\subsection{Relativistic Transparency}

If we consider a plasma place between the two circularly-polarised waves, then for laser intensities such that $a_0 \gg 1$, the motion of electrons becomes sufficiently relativistic that their average Lorentz factor $\bar{\gamma} \approx \left( 1 + a_0^2 \right)^{1/2}$ (neglecting radiation reaction -- discussed below) must be accounted for in the effective plasma frequency $\omega_{p}/\sqrt{\bar{\gamma}}$. Since $\bar{\gamma} \gg 1$, this means the plasma frequency is dramatically reduced relative to the laser frequency and a plasma which was overdense at lower intensity becomes underdense, allowing the laser to propagate through the plasma. As the plasma frequency is related to the plasma density by $\omega_{p} \propto n^{1/2}_{e}$, and given that the non-relativistic critical density is $n_{C0}=m_e \varepsilon_0 \omega_L^2/e^2$, the relativistically corrected critical density becomes $n_C=\bar{\gamma} n_{C0}$. At the intensities considered in this work, $a_0 \gg 1$ and so $\bar{\gamma} \approx a_0$ and using the value for $a_0$ given above the relativistically-corrected critical density is $n_C \approx 600n_{C0} (I_{24}\lambda^2_{\mu m})^{1/2}$.

A further correction is due to the strong radiation reaction experienced by electrons at intensities sufficient for quantum effects to become apparent, thus requiring the introduction of a damping correction which reduces $\bar{\gamma}$ below $a_0$ \cite{Zhang15a}. When radiation damping is weak the critical density is the same as the undamped value given above, that is $n_{C}^{W}$. However, for intensities $I_{24} \gtrsim 1$, damping becomes strong and the critical density becomes $n_C^{S} \approx 2a_0n_{C0}$. Here we use the undamped classical value $n_C=n_C^W=a_0 n_{C0}$ for simplicity, and here on when discussing under/overdense plasma we mean relativistically under/overdense.

\subsection{\label{StrongFields} Strong-field effects}

The important strong-field QED effects in laser matter interactions have been discussed extensively in the literature (see, for example \cite{Kirk09a,Ridgers14a,Gonoskov15}), we review them here for convenience.  The characteristic field of non-linear quantum electrodynamics is the Sauter-Schwinger field $E_S=m_e^2 c^3/e\hbar\approx1.32\times{10}^{18} \, \mathrm{Vm}^{-1}$. For electrons and positrons, the importance of QED effects is primarily governed by the dimensionless, Lorentz-invariant parameter

\begin{equation} \label{eta}
\eta \equiv \frac{e\hbar}{m_e^3 c^4} \vert F_{\mu\nu} p^{\nu} \vert = \frac{E_{RF}}{E_s} \approx \frac{\gamma}{E_s} \vert \mathbf{E}_{\perp}+\mathbf{v}\times\mathbf{B} \vert 
\end{equation}

\noindent where the last equality is valid when the electron (positron) is ultra-relativistic.  $p^{\mu}$ is the four-momentum of an electron travelling in a background electromagnetic field with field tensor $F^{\mu\nu}$, $\gamma$ is the Lorentz factor of an electron travelling at velocity $\mathbf{v}$, $\mathbf{E}_{\perp}$ is the component of the electric field perpendicular to the electrons motion (i.e. perpendicular to $\mathbf{v}$) and $\mathbf{B}$ is the magnetic field.

An electron at a magnetic node of the standing wave formed by counter-propagating circularly-polarised laser pulses performs circular motion with the centripetal force provided by the component of the laser's electric field perpendicular to its motion $\mathbf{E}_\perp$. In this case $\eta=\gamma E_{\perp}/E_S$. At high intensities the average value for the Lorentz factor of an electron $\bar{\gamma}=(1+a_0^2)^{1/2}\approx a_0$ (again neglecting radiation reaction -- see Refs. \cite{BellKirk08a,Zhang15a} for the equivalent discussion including radiation reaction), meaning the average $\eta$ value $\bar{\eta} \approx a_0 E_{\perp}/E_S \approx 1.75 I_{24} \lambda_{\mu m}$ for counter-propagating circularly-polarised beams (assuming $E_{\perp} = 2E_0$). For pair production to become important we require that the electromagnetic field strength approaches the Sauter-Schwinger field, i.e. $\bar{\eta} \sim 1$, from which it can be seen that for counter-propagating lasers of $\lambda_{\mu m}=1$ we require the intensity be $I_{24} \sim 0.57$.

For $\eta \sim 1$, three quantum effects predominantly affect the behaviour of electrons, positrons, and $\gamma$-ray photons interacting with an intense laser. These are non-linear Compton scattering and pair production via the Trident and multi-photon Breit-Wheeler processes.  The first, non-linear Compton Scattering, is the scattering of $n$ laser photons $\hbar\omega_L$ by an electron resulting in a single high-energy $\gamma$-ray photon $\hbar\omega_{\gamma}$, i.e. $e^-+n\hbar\omega_L\to\hbar\omega_{\gamma}$. It governs the emission of high-energy $\gamma$-ray photons by an electron or positron accelerated by the laser fields. The average energy of the emitted gamma-ray photon is $(\hbar\omega_\gamma)_{av}\approx0.44\eta$ times the emitting electrons energy \cite{Kirk09a} and so for $\eta \sim 1$ each emission leads to a large change in the electrons energy and the electron's motion becomes stochastic \cite{Duclous11a, Shen71a}. However, it has recently been shown that a modified-classical approach to radiation reaction using the ultra-relativistic form of the Landau Lifshitz equation \cite{Landau75a} including the Gaunt factor $g(\eta)$ for synchrotron emission \cite{Baier91a} describes the average motion of the electron population to good approximation \cite{Ridgers17a,Niel18}.

The trident process occurs when a virtual photon decays into an electron-positron pair which is subsequently separated by an external electromagnetic field. The rate of this process increases relatively slowly with intensity \cite{BellKirk08a} and as such it is typically ignored as it will be here.

The final process of multi-photon Breit-Wheeler pair production is similar to the Trident process but results from a real photon, rather than virtual, interacting with laser photons to produce a pair, i.e. $\hbar\omega_{\gamma} + n\hbar\omega_L\to e^- + e^+$. This process is dependent on a second Lorentz-invariant quantum parameter for the photon $\hbar\omega_{\gamma}$,

\begin{equation} \label{chi}
\chi \equiv \frac{e\hbar^2}{2m_e^3 c^4} \vert F_{\mu\nu} k^{\nu} \vert = \frac{\hbar\omega_{\gamma}}{2m_ec^2}\vert \mathbf{E}_{\perp}+c\hat{\mathbf{k}}\times\mathbf{B}\vert
\end{equation}

\noindent where $\hbar k^{\nu}$ is the 4-momentum of the photon interacting with a background (laser) field, $\hbar\omega_{\gamma}$ is its energy and $\mathbf{k}$ is its 3-wavevector.  In the case of photons emitted by an electron performing circular motion at the magnetic node, the average value of $\chi$ is given by $\bar{\chi}\approx [(\hbar\omega_{\gamma})_{av}/2m_ec^2] (E_{\perp}/E_S$).

Rates for non-linear Compton scattering and multi-photon Breit-Wheeler pair production are known and are conveniently reviewed in, for example, \cite{Ritus85a}. These rates were calculated under the assumptions of a quasi-static and weak external field. The first requires that the formation length of processes be short relative to the characteristic length-scale of change in the external field so that the rates may be calculated for a constant field. In general the rates depend not only on $\eta$ and $\chi$ (as defined in equations (\ref{eta}) and (\ref{chi})) but also the parameters $F=\vert E^2-c^2B^2 \vert/E_S^2$ and $G=\vert \mathbf{E} \cdot c\mathbf{B} \vert/E_S^2$. However, provided the weak-field approximation applies, i.e. that $E_0 \ll E_S$ ($F,G \ll 1$ and $\eta^2 \gg \max{(F,G)}$), the rates can be treated using the constant crossed-field configuration (as a function of $\eta,\chi$ only). In this case the rate that an electron with energy $\gamma mc^2$ emits a photon and a photon with energy $\hbar\omega_{\gamma}$ decays to an electron-positron pair are \cite{Ritus85a,Bashmakov14a}
\begin{equation}
W_{\gamma} =  \frac{\sqrt{3}\alpha_f c}{\lambda_c}\frac{\eta}{\gamma} \int_0^{\eta/2} d\chi \frac{F(\eta,\chi)}{\chi} \quad\quad
W_{\pm} = \frac{2\pi\alpha_f c}{\lambda_c}\frac{m_ec^2}{\hbar\omega_{\gamma}}\chi T_{\pm}(\chi).
\end{equation}
\noindent $\lambda_c$ and $\alpha_f$ are the Compton wavelength and fine-structure constant.  $F(\eta,\chi)$ is the quantum synchrotron function (whose form is given in \cite{Ridgers14a}) and $T_{\pm}\approx 0.16 K_{1/3}[2/(3\chi)]/\chi$ ($K_{1/3}$ is a Bessel function of the second kind). When developing the semi-analytical model we will assume that $\eta=\bar{\eta}$ \& $\chi=\bar{\chi}$ in these equations for the rates.

\subsection{Identifying the cascade regimes}
\label{section:cascade_regimes}

We now have the formulae required to develop our semi-analytical model of dense pair plasma production.  Laser-induced electron-positron cascades have been previously investigated for various laser intensities, targets and laser pulse shapes (e.g. \cite{BellKirk08a,Kirk09a,Elkina11a,Grismayer17a}). The dynamics of cascades are complicated and in general analytical solutions are unattainable. Recently Grismayer et al. \cite{Grismayer17a} derived semi-analytical scalings for the growth of the cascade from a small number of seed electrons. This has been extended to include cascades from an initially underdense plasma by Luo et al. \cite{Luo18} and Del Sorbo et. al \cite{DelSorbo18}. The latter cases, where the seed is an initially present electron-ion plasma, is far more likely to be realised in experiments.  We extend the analysis of this case to include the case where this electron-ion plasma has density close to or above the critical density.

We can write the coupled rate equations for the number of pairs $N_\pm$ and the number of photons $N_{\gamma}$ \cite{Bashmakov14a}:
\begin{equation}
\dot{N}_\pm=W_\pm N_{\gamma},
\end{equation}
\begin{equation}
\dot{N}_{\gamma}=2W_{\gamma}N_\pm - W_{\pm}N_{\gamma}.
\end{equation}

\noindent Here we have neglected photon emission from the electrons in the initially present electron-positron plasma.  Recently it has been shown that the generated electron-positron plasma radiates more energy \cite{DelSorbo18}.

These equations have solutions of the form $N_{\pm,\gamma}(t) \propto \exp{\Gamma t}$, where the cascade growth rate is
\begin{equation}
\Gamma=\frac{W_\pm}{2}\left( -1 + \sqrt{1+\frac{8W_{\gamma}}{W_\pm}} \right).
\end{equation}

If the time between emissions is small, i.e. if $W^{-1}_{\gamma,\pm} \ll \omega_L^{-1}$, then the distance a particle can travel from its parent before emitting itself is short ($\ll \lambda$). Hence, if the initial density of electrons in the field region is $n_0$, the density of electron-positron pairs will evolve in a manner similar to $N_\pm$ or according to
\begin{equation}
\label{number_density_underdense}
n_\pm = n_0 \left( \exp{\left( \Gamma t \right)} - 1 \right).
\end{equation}
This assumes that the plasma formed of the original electron ion plasma and the generated pair plasma is everywhere underdense (i.e. the plasma frequency of the plasma is less than the laser frequency) so that the lasers are perfectly transmitted and the standing wave may form without disturbance.

The previous assumption breaks down if the number density of electrons in the original electron-ion plasma or the self-generated pair plasma approaches the relativistically-corrected critical density for the laser light $n_C = \bar{\gamma}n_{C0}=\bar{\gamma} m_e \varepsilon_0 \omega^2_L/e^2$. In this case the plasma can shield the electrons and positrons from the laser fields by the skin effect and pair production is curtailed \cite{Ridgers12a,Ridgers13a}. To account for the skin effect we propose the following heuristic modification to equation (\ref{number_density_underdense}) (which we will see works well when compared to simulation results):
\begin{equation}
\label{number_density_overdense}
n_\pm = n_0 \left( \exp{\left( \Gamma t \right)} - 1 \right) \exp{\left( - \frac{\lambda}{\delta_S} \right)}.
\end{equation}
The factor $\exp{(-\lambda/\delta_S)}$ for constant wavelength $\lambda$ and relativistic skin depth $\delta_S = \sqrt{\bar{\gamma}} c/\omega_{p} \in (0,\infty)$ -- $\omega_p$ is the electron plasma frequency in the originally present electron-ion plasma. That is, in the limit that the relativistic skin depth $\delta_S$ is large (i.e. $\delta_S \gg \lambda$) the exponential factor goes over to unity, giving the original underdense case in equation (\ref{number_density_underdense}), and as the skin depth becomes small ($\delta_S \ll \lambda$) the number of pairs approaches zero as expected due to the reduced interaction volume of the strong laser-fields with the plasma. 

As the cascade progresses the number density of pairs can continue to grow until either it reaches the relativistically-corrected critical density $n_C=\bar{\gamma}n_{C0}=\bar{\gamma} m_e \varepsilon_0 \omega_L^2/e^2$, at which point the laser energy is fully absorbed by the pair plasma \cite{Zhang15a,Grismayer17a}, depleting the field -- the cascade saturates. We can estimate the saturation time ($t_C$) by setting $n_\pm=n_C$ in equation (\ref{number_density_overdense}), which gives

\begin{equation}
\label{saturation_time}
	t_C = \frac{1}{\Gamma}\ln{\left( \frac{n_C}{n_{0}}\exp{\left( \frac{\lambda}{\delta_S} \right)} + 1 \right)},
\end{equation}

\noindent provided the laser pulse length $\tau_P$ is longer than $t_C$ the cascade can saturate resulting in the formation of a relativistic-critical density pair plasma.  

Using equation (\ref{number_density_overdense}) for $n_{\pm}$ and equation (\ref{saturation_time}) for $t_C$, we can discern three regimes for the cascade, depending on the initial target density and the laser intensity:

\begin{enumerate}

\item \emph{Exponential growth}: if the laser intensity is sufficiently high and the plasma is relativistically underdense, i.e. the skin-depth is larger than the laser wavelength, then the interaction volume of the laser with the plasma is large enough that a cascade can be initiated, generating a pair plasma whose density grows exponentially according to (\ref{number_density_overdense}).

\item \emph{Saturation}: $t_C$ is less than the laser pulse duration $\tau_p$ and the cascade develops until a critical density pair plasma forms and the cascade saturates. Once this occurs the plasma partially absorbs (and partially reflects) the remainder of the laser pulse.  Therefore for $\tau_p>t_C$ we would expect appreciable laser absorption caused by non-linear Compton scattering and the resulting radiation reaction in the generated critical density pair plasma \cite{Zhang15a,Grismayer17a,DelSorbo18}.  If $\tau_p<\tau_C$ the cascade remains in the exponential growth phase, so the density of pairs remains low and thus there will not be appreciable laser absorption.

\item \emph{Cascade suppressed}: if the electron number density in the initially present electron-ion plasma is higher than the relativistically-corrected critical density the laser-plasma interaction region is severely limited by the small skin depth. In this case equation (\ref{number_density_overdense}) predicts a progressively lower number density of pairs produced scaling inversely with the target density, i.e. the cascade is suppressed even if the laser intensity high enough to cause a cascade in an initially underdense plasma.

\end{enumerate}

\section{Verifying the cascade regimes with PIC simulations}
\subsection{Simulation set-up} \label{sim_set_up_1}

To investigate the semi-analytical scaling for pair plasma number density given in equation (\ref{number_density_overdense}) and the cascade regimes predicted in section \ref{section:cascade_regimes}, we performed 1D and 2D PIC simulations using the QED-PIC code \textsc{epoch} \cite{Arber15a}.  \textsc{epoch} includes strong-field QED processes using a now standard Monte-Carlo model, described in Ref. \cite{Ridgers14a}. We simulated the specific case of two counter-propagating circularly-polarised lasers interacting with a $l=1\, \mathrm{\mu m}$ thick hydrogen plasma target. The peak intensity $I$ (of one of the counter-propagating laser pulses) and the initial electron number density in the target $n_0$ were varied between $I_{24}=I/(10^{24}$\ Wcm$^{-2})$ $\in[0.1,5]$ and $n_0\in[10^{26},10^{31}]$m$^{-3}$, respectively.

Each laser had a continuous flat temporal-profile, with a Gaussian ramp-up (with time-scale $\lambda/2c$), in order to reduce numerical artefacts due to discontinuities. The laser wavelength was chosen to be $\lambda=1 \, \mathrm{\mu m}$ to be close to expected values from future multi-PW laser facilities. In the 2D simulations, the beam was given a Gaussian profile in the transverse direction, with focal spot size $2.5 \, \mathrm{\mu m}$.

For simplicity, the target was a fully-ionised Hydrogen plasma.  The target was initialised with density given by a top-hat profile in the $x$-direction (the direction of laser propagation), so that $n(x,t=0)=n_0$ for $\vert x\vert\leq 0.5 \, \mathrm{\mu m}$ and zero elsewhere.  In the 2D simulations, the target was simply extended infinitely in the transverse ($y$) direction to form a foil.

In the 1D simulations the spatial domain was $8 \, \mathrm{\mu m}$ in length and discretised with $1024$ cells. A reflecting boundary was placed at $x=0$ in order to reduce computational load (this is reasonable when considering the longitudinal symmetry of the physical set-up). A total of $100\times1024\approx10^5$ macroparticles were used to initialise the target plasma, split equally between protons and electrons. While in the 2D simulations a domain size of $6 \, \mathrm{\mu m} \times 12 \, \mathrm{\mu m}$ was used, discretised with 600 cells in the $x$-direction and 1200 cells in the $y$-direction. The target was represented by $64\times600\times1200\approx4.6\times10^7$ macroparticles again split evenly between electrons and protons.

Examination of the evolution of pair and plasma densities as well as the laser absorption for the case $I_{24}=1$ and varying $n_0$ showed satisfactory convergence for the above number of cells and macroparticles. Further increases in spatial resolution or particles per cell produced no substantial changes to the generation of a dense pair plasma, and would have been prohibitively computationally expensive for $I_{24}>1$.  We have neglected collisions.

\subsection{1D Simulation results}

\begin{figure}
  \includegraphics[width=\linewidth]{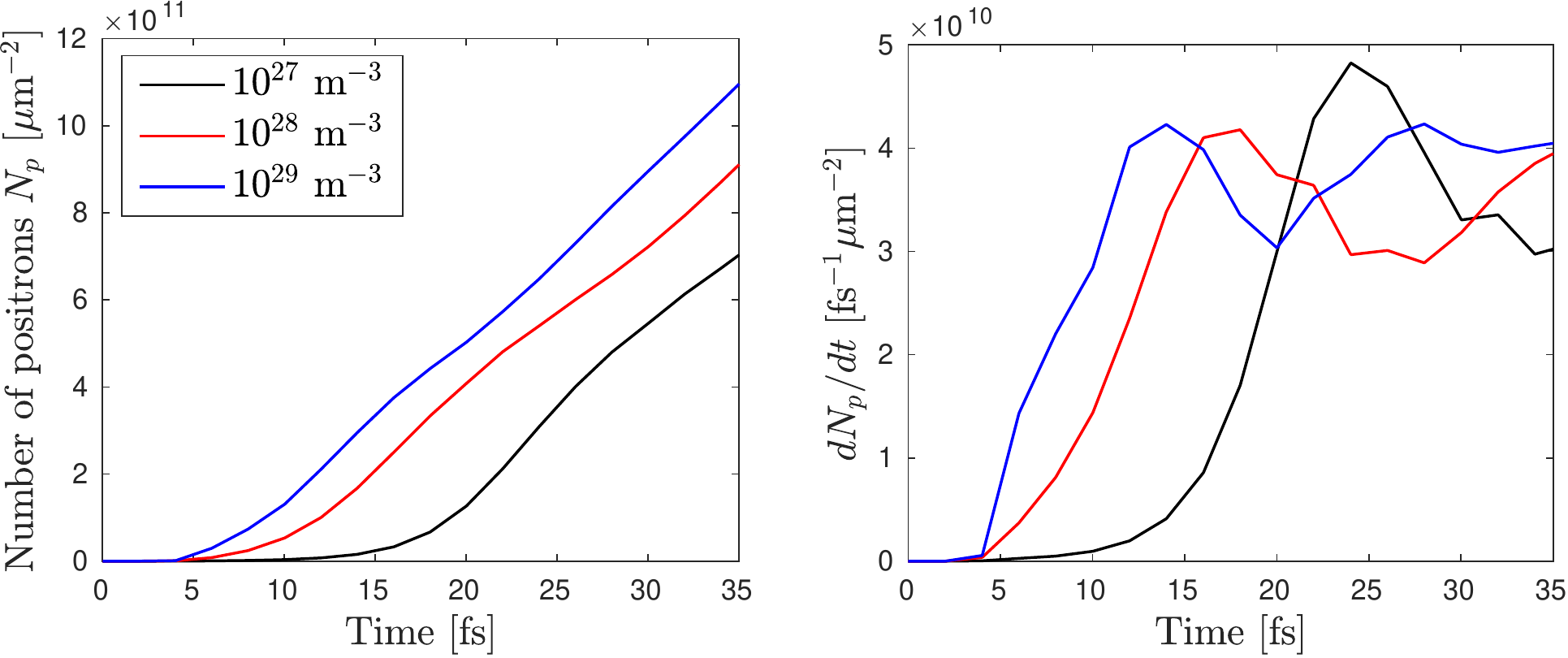}
	\caption{\label{PairsRates} \small (Left) The number of positrons $N_{p}$ generated in a cascade for $I_{24}=1$ and $n_0=10^{27},10^{28}\  \&\  10^{29} \ \rm m^{-3}$ (corresponding to the colours denoted by the legend). (Right) Positron production rate $dN_p/dt$.}
\end{figure}

Figure \ref{PairsRates} shows the time evolution of the number of pairs $N_{\pm}$ and the production rate $\dot{N}_{\pm}$ for selected 1D simulations. As previously seen by Grismayer et al. \cite{Grismayer17a} two distinct phases of the cascade are observed: (1) an exponential growth phase where the plasma remains sufficiently underdense that the standing wave pattern is not disturbed; (2) a saturation phase where the pair plasma density reaches the relativistic critical density and the laser energy is absorbed and the rate of pair production levels off. The dynamics of the cascade is as follows: at $t=0$, the laser hits and begins to bore through the target, compressing the electron density. In the cases considered in figure \ref{PairsRates}, the target is initially underdense and the laser propagates mostly unhindered by the plasma, forming a standing wave in the plasma volume after a time $l/c \approx 3.33 \, \mathrm{fs}$. If the laser intensity is such that $\eta \sim 1$, then $\gamma$-ray photons radiated by the electrons (and positrons) can decay to form electron-positron pairs within the laser-plasma interaction volume, initiating the cascade.

From figure \ref{PairsRates}, we note that increasing the target density does not change the overall behaviour of the cascade in the exponential growth phase; that is, the functional dependence of the number of pairs produced and the production rates on time remain similar as we vary the target density, but are offset temporally as the cascade must develop from fewer seed electrons in a lower density target.

\begin{figure}
  \includegraphics[scale=0.8]{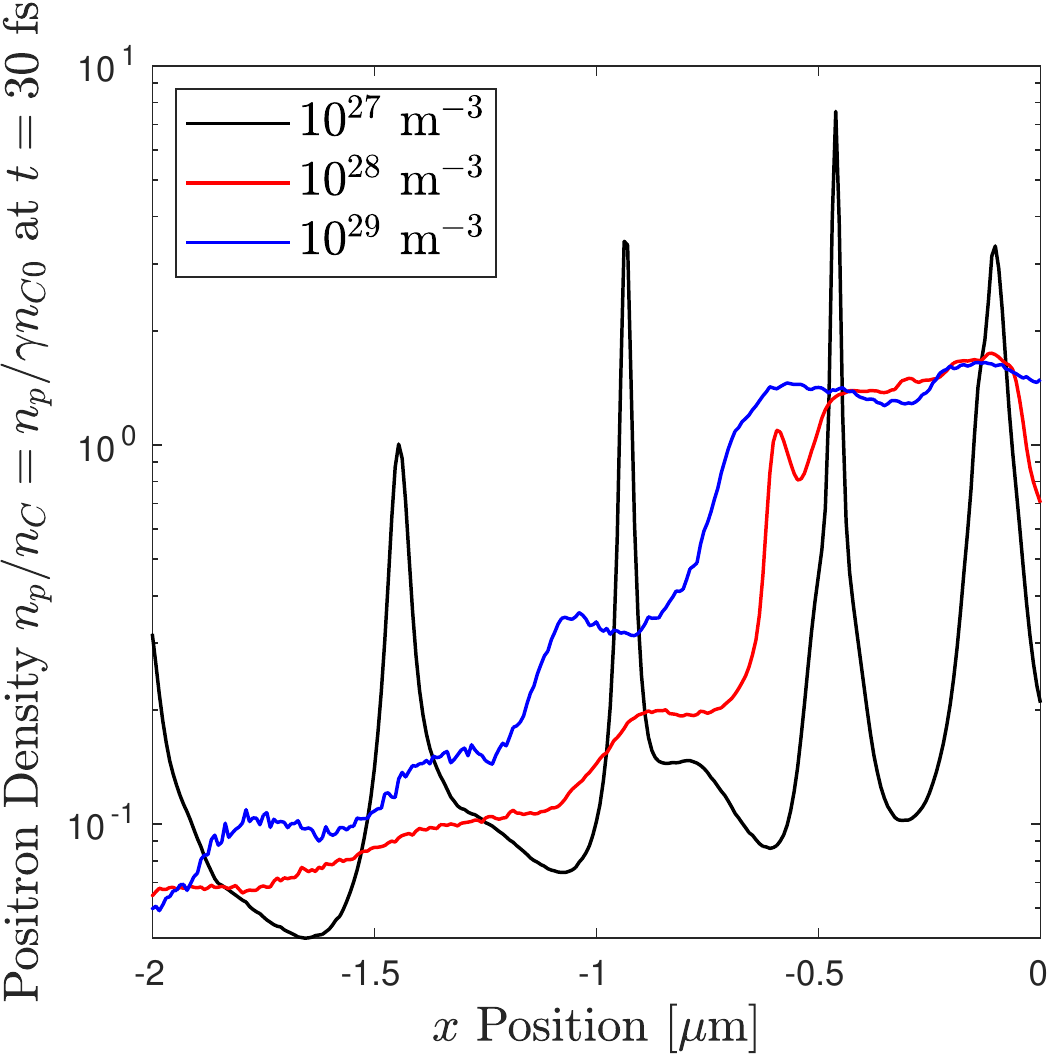}
	\caption{\label{PairsDensity} \small Positron densities at time $t=30 \ \rm fs$ for the same parameters as the those in figure \ref{PairsRates}.}
\end{figure}

Figure \ref{PairsDensity} shows the density of positrons produced at $t=30$\ fs.  This supports the description of the temporal evolution of the cascade given above.  For the lowest target density the cascade has yet to saturate and the standing wave in the electromagnetic fields has yet to be disrupted by the generated pair plasma.  As a result the created positrons congregate in the nodes of electric field (the magnetic nodes are unconditionally unstable).  For higher density targets the cascade has saturated at the critical density, disrupting the standing wave and so the periodicity in positron density is lost.

\subsubsection{Pair plasma generation for varying laser intensity and target density}

\begin{figure}
	\includegraphics[width=0.8\linewidth]{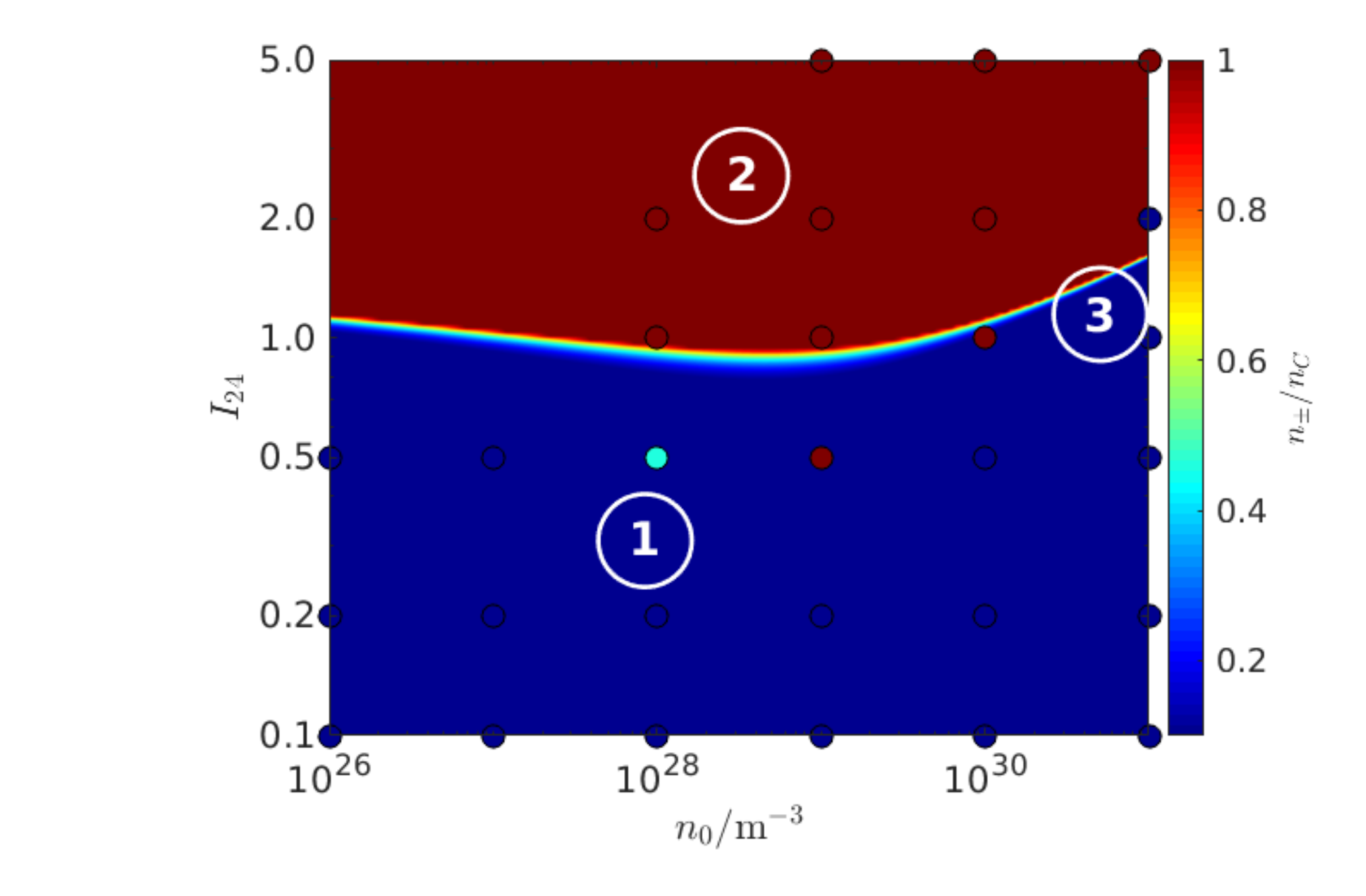}
	\caption{\label{Pairs} \small The normalised pair plasma density $n_\pm/n_C=n_\pm/\gamma n_{C0}$ from equation (\ref{number_density_overdense}) at time $t=40 \rm \ fs$ (capped at $n_\pm /n_C=1$).  Dots are the results of 1D simulations with the specified parameters (the pair density is capped at the critical density).  Illustrated are the cascading regimes 1 -- exponential growth, 2 -- saturation \& 3 -- cascade suppressed, identified in section \ref{section:cascade_regimes}.}
\end{figure}

Figure \ref{Pairs} shows the maximum, over the spatial domain, of $n_\pm / n_C$ at time $t=40 \ \rm fs$  \footnote{This can be above the critical density even before saturation has occurred (see the density spikes in figure \ref{PairsDensity}), potentially leading to the spurious identification of a critical density pair plasma in figure \ref{Pairs}.  However, this is only a problem late in the exponential growth phase, which is reached after 40\ fs for a narrow range of initial densities and laser intensities.} as a function of the intensity of each laser pulse and the initial electron number density in the target. For both the predictions of the semi-analytical results presented in section \ref{section:cascade_regimes} and the simulations, the density is capped at $n_C$ to show the different regimes clearly. The model predictions are shown by the colour plot and the 1D simulation results by the coloured dots. The theory and simulation results are qualitatively well-matched, we can see the demarcation of the regimes predicted by the model are borne out by the simulation results. Both the model and the simulations predict that the region where a dense pair plasma is created extends to lowest intensity when the initial density of the target is close to the critical density.  In the simulations a dense pair plasma forms at intensity $>5\times10^{23} \, \mathrm{Wcm}^{-2}$.  This is in line with Zhu et al. \cite{Zhu16a}, who observe a similar effect using structured plasma to strongly focus two $\sim 10^{22} \ \rm Wcm^{-2}$ pulses (to $>10^{23}$\ Wcm$^{-2}$) incident on a near-critical hydrogen plasma.  Note that there is some discrepancy between the simulations and the simple model when the target density close the critical density.  At this density we expect very complex plasma behaviour, not easy to capture in a simple scaling law, and so would not expect perfect agreement.

\subsubsection{Absorption due to pair generation}

We now consider the impact of the generated dense pair plasma on laser absorption. As described in \S \ref{StrongFields}, the processes of non-linear Compton scattering and multi-photon Breit-Wheeler pair production both result in the absorption of photons from the background field. In addition energy can be absorbed from the laser pulse as the laser-fields accelerate the generated pairs against the radiation reaction force.  The latter classical absorption dominates over the former quantum effect \cite{Seipt17a} which is neglected in the simulations.

A model for absorption has been derived by Grismayer et al.  \cite{Grismayer16a} and developed for dense targets by Del Sorbo \cite{DelSorbo18}. In this model it is assumed that strong absorption will occur once the pair plasma density reaches the critical density. The time for this to occur is given by equation (\ref{saturation_time}) above, which results in a new adaptation of the model to include the skin effect in dense targets. The laser absorption is then the ratio of the energy absorbed to the laser energy which, for $t_C \leq \tau_P$, is
\begin{equation}\label{Absorption}
	\frac{\mathcal{E}_a}{\mathcal{E}_L}=\left( 1-\frac{t_C}{\tau_P} \right) \Theta(\tau_P - t_C)
\end{equation}
where the Heaviside function $\Theta(\tau_P - t_C)$ accounts for the model assumption that the absorption is negligible, i.e. zero, for $t_C > \tau_P$. For continuous beams, as considered in this paper, the pulse duration is taken to be the time from when the laser pulses collide to the time at which the absorption is measured.

\begin{figure}
	\includegraphics[width=0.8\linewidth]{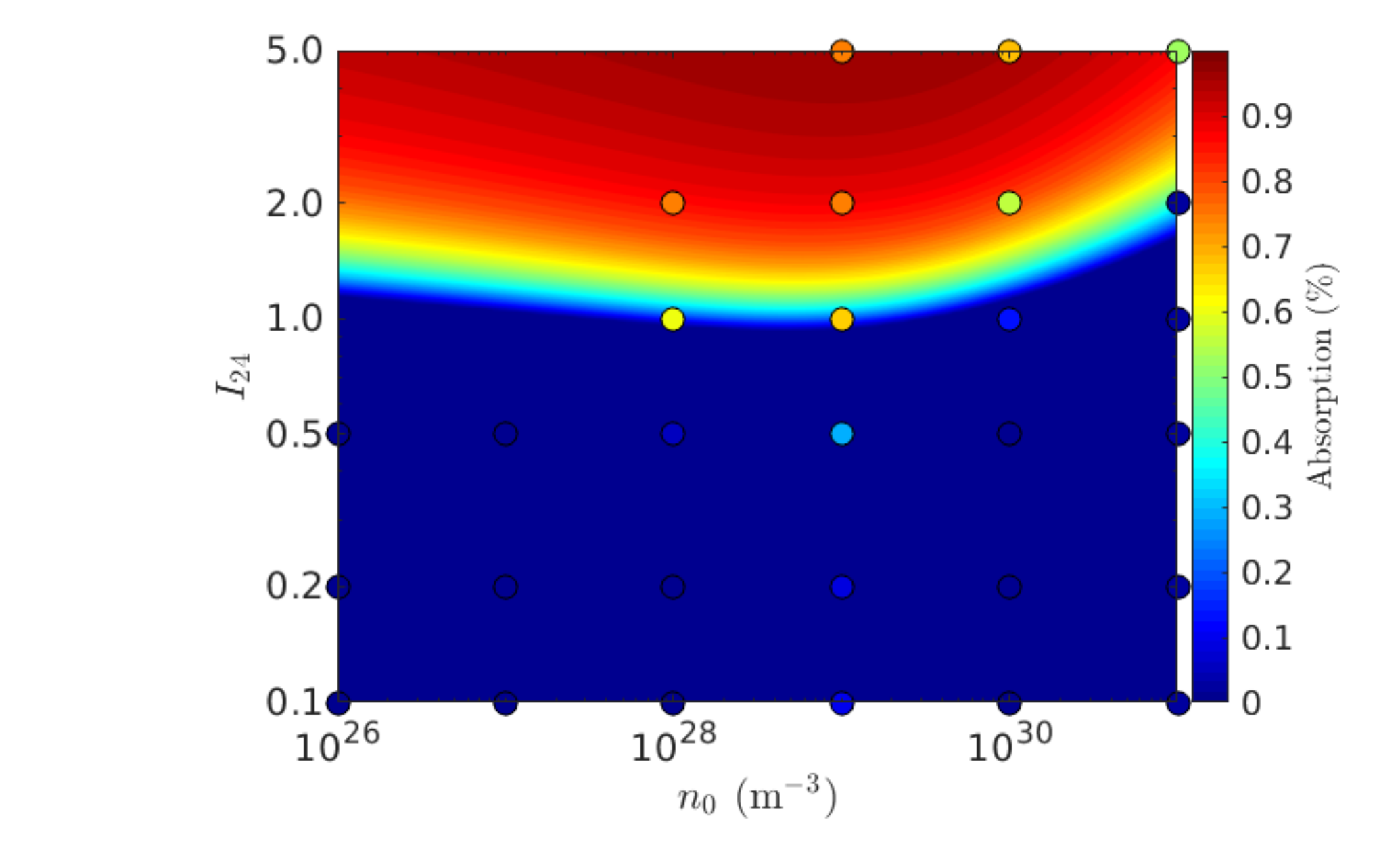}
	\caption{\label{Absorb} \small Predicted absorption fraction of the laser energy (colour scale) and 1D simulation results (dots) at $t=42 \, \rm fs$, highlighting the impact of the generation of a dense pair plasma.}
\end{figure}

Figure \ref{Absorb} shows the percentage of laser energy absorbed as predicted by the above model, i.e. equation (\ref{Absorption}), and obtained from the 1D simulations $42 \, \rm fs$ after the laser hits the target. Again the simple model predicts the simulations results qualitatively very well, although discrepancies at around the critical density are again seen.  The agreement with the simple model and a comparison with figure \ref{Pairs}, demonstrates that strong laser absorption is strongly correlated with the generation of a dense pair plasma.  Recently it has been shown that the dense pair plasma generated in cascades radiates the laser energy effectively and it is this which results in the laser absorption and thus ignoring absorption until the onset of a cascade is justified \cite{Grismayer17a,DelSorbo18}.  However, absorption caused by the electrons in the original near critical plasma may be a cause of the enhanced absorption in the simulations at the critical density when compared to the model \cite{Zhang15a} (in addition to enhanced absorption due to more pair creation in the simulations of near-critical targets than the model predicts). 

\begin{figure}
	\includegraphics[width=\linewidth]{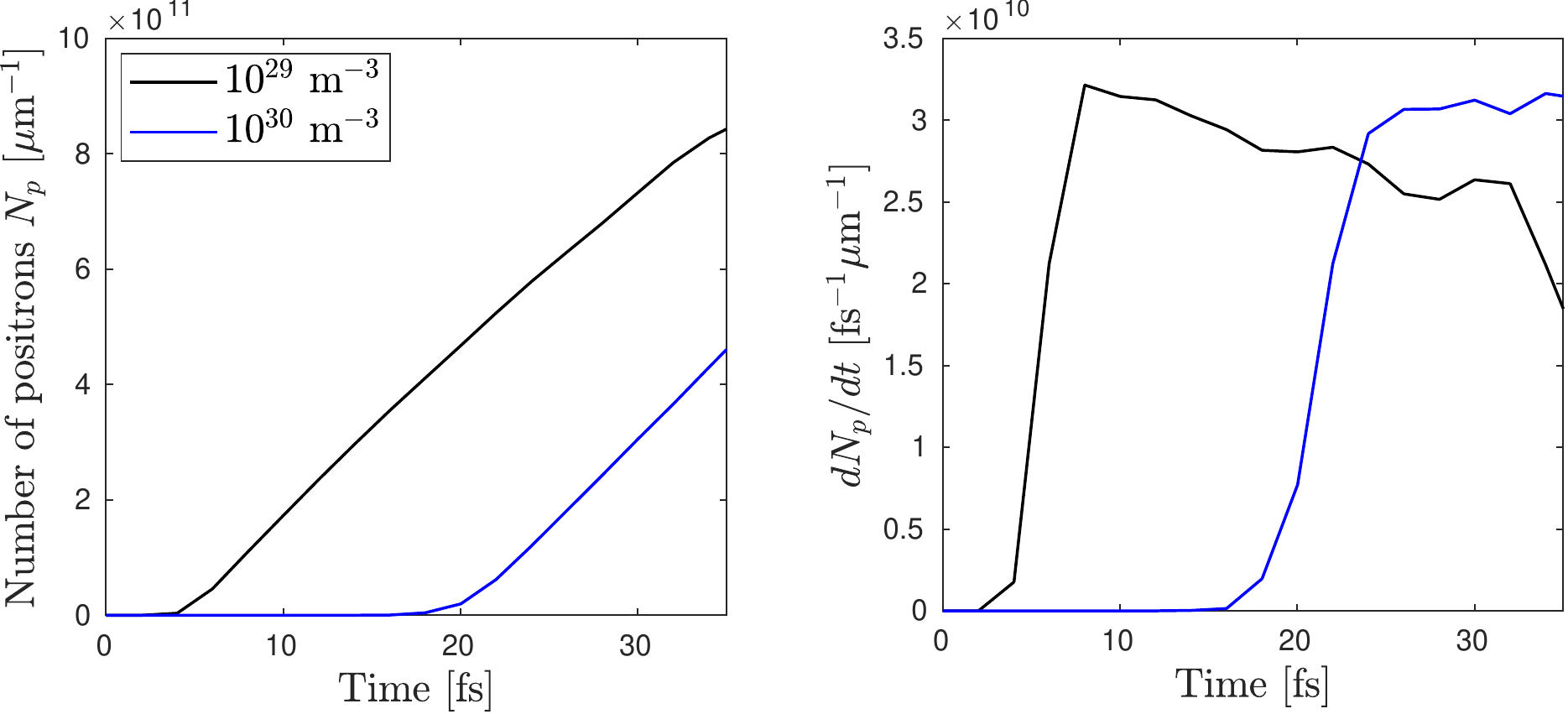}
	\caption{\label{2DPairRate} \small 2D simulations results showing: (left) the number of positrons $N_{p}$ generated in a cascade for $I_{24}=1$ and $n_0=10^{29}\ \&\ 10^{30} \ \rm m^{-3}$ (corresponding to the colours denoted by the legend). (Right) Positron production rate $dN_p/dt$. The behaviour is qualitatively the same as in the 1D simulations, the results of which were shown in figure \ref{PairsRates}.}
\end{figure}

\subsection{2D simulation results}

In order to test the robustness of the identified cascading regimes in a more realistic scenario, 2D simulations were performed (with parameters described in section \ref{sim_set_up_1}).  Figure \ref{2DPairRate} shows that, as in the 1D case, the cascade exhibits an exponential growth phase followed by saturation.

\begin{figure}
	\includegraphics[width=1.1\linewidth]{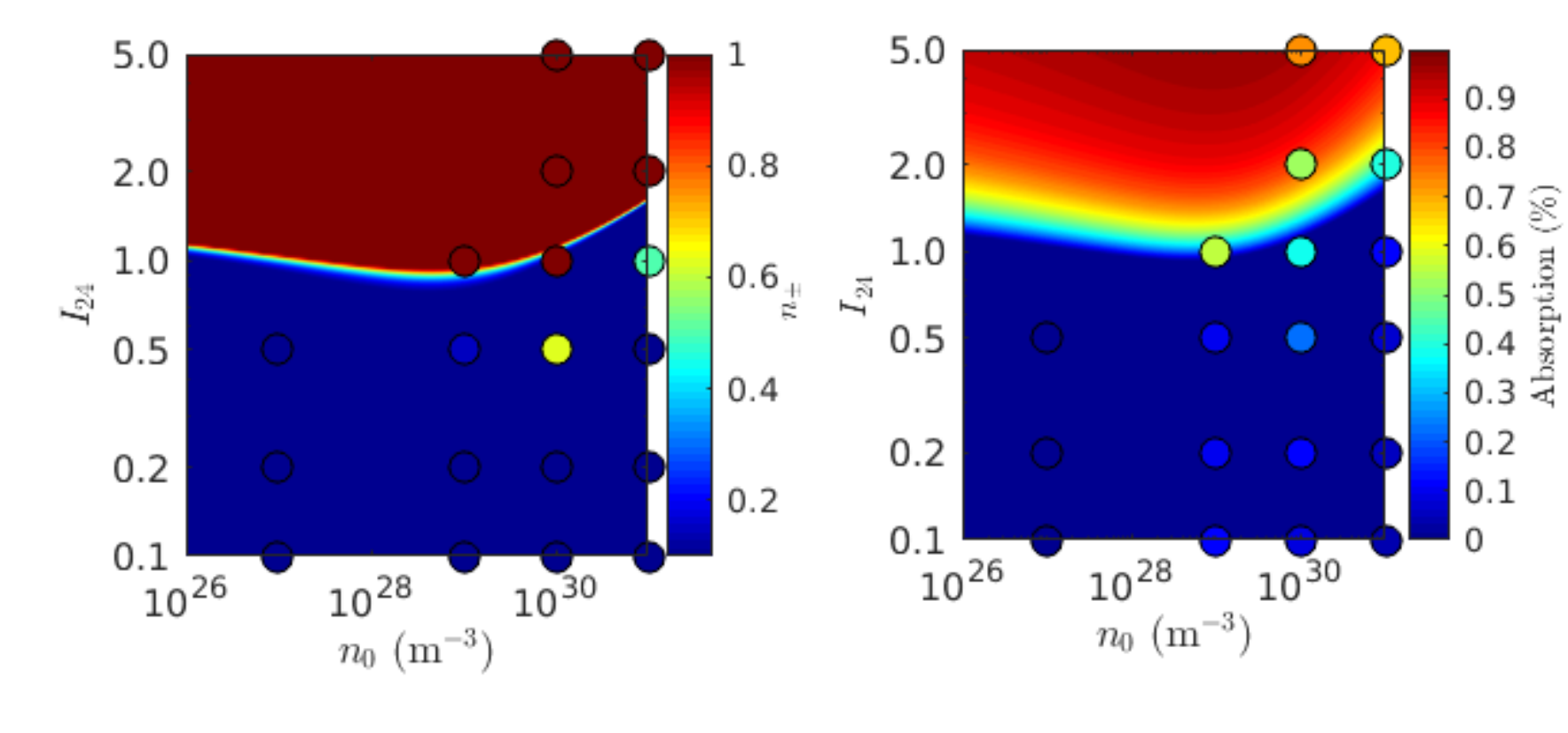}
	\caption{\label{2DPairs_and_abso} \small \centering The normalised pair plasma density $n_\pm/n_C=n_\pm/\gamma n_{C0}$ from equation (\ref{number_density_overdense}) at time $t=40 \rm \ fs$ (capped at $n_\pm /n_C=1$).  Dots are the results of 2D simulations with the specified parameters (the pair density is capped at the critical density).}
\end{figure}

The qualitative similarity of the cascade development between the 1D and 2D simulations, as demonstrated by figures \ref{PairsRates} and \ref{2DPairRate} suggests that the simple model presented in section \ref{section:cascade_regimes} may also work well for describing the pair density and absorption in 2D simulations.  In figure \ref{2DPairs_and_abso} we see the 2D equivalents to figures \ref{Pairs} \& \ref{Absorb}, i.e. a comparison to the model predictions for average pair density and laser absorption.  Qualitative agreement is again seen between the model and the simulations, demonstrating the usefulness of the model in more experimentally applicable 2D simulations.

\section{Discussion}

We have investigated the case of two counter-propagating circularly-polarised lasers interacting with a thin fully-ionised hydrogen plasma. Electron-positron cascades in these interactions can lead to the formation of dense pair plasmas dependent on the target density and laser intensity.  We have identified several regimes of laser-plasma interaction based on whether or not a cascade can develop.  We have shown, by developing a simple model that for intensities $I_{24}>1$ the growth rate in the number of pairs is sufficient for the production of a relativistic-critical density pair plasma in a time comparable to a typical laser-pulse duration $\sim 40\ $fs.  The cascade saturates and the produced pair plasma absorbs a substantial fraction of the laser energy.  The optimum initial hydrogen target density for this to occur is close to the relativistic critical density with the intensity required to develop a cascade to saturation increasing as the density decreases (figure \ref{Pairs}). We showed in figure \ref{PairsRates} that as the initial target density decreases below the relativistically-corrected critical density the temporal evolution of the number of pairs looks similar (exponential growth followed by saturation) but is shifted to later time as the cascade must grow from fewer particles and the growth rate per particle is the same. Thus for a given intensity the production of a dense pair plasma depends simply on whether the initial density of the target is sufficient for the density of pairs to reach the relativistic critical density in time $t_C$ given by equation (\ref{saturation_time}) which must be less than the laser pulse duration.  Most previous studies have considered cascades seeded by targets of much lower density \cite{Nerush11,Grismayer17a}, in which case a higher intensity is required to initiate the cascade.  In addition many previous studies considered linearly-polarised lasers which are favourable to cascades but make comparison to simplified analytical theory more difficult, suggesting that we may be underestimating the reduction in the intensity required to initiate a cascade by the use of near-critical density targets.

  As the initial target density increases beyond the relativistic critical density pair cascades are rapidly suppressed due to the shielding of the laser fields by the skin effect in the dense initially present electron-ion plasma and the laser pulses are reflected.  We found that a hard cut-off in the number of pairs produced at the relativistic critical density does not match the simulations as well as a heuristically introduced exponential fall-off proportional to $\exp(-\lambda/\delta_c)$.  We have seen that the inclusion of a transverse direction in the simulations reduces the generation of pairs (as seen for tighter focusing \cite{Jirka18}), although near-critical targets are still optimum for pair plasma production, as has previously been shown for gamma-ray emission \cite{Nakamura12,Brady12,Brady13b}. This reduction can be attributed to transverse spreading of pairs from laser focus due to ponderomotive and thermal pressure.

  Recent work has shown that photon polarisation \cite{King13} and electron spin \cite{DelSorbo17,DelSorbo18b,Seipt18} could affect the development of the cascade.  These effects are not considered here.  These effects should change the cascade growth rate and so may affect the intensity required for saturation. The fact that cascade saturation and suppression depend entirely on plasma processes suggests that these electron spin and photon polarisation processes will not change the various phases of the cascade identified in section \ref{section:cascade_regimes} and so will not qualitatively change the cascade regimes presented here. In addition we only consider a very simple counter-propagating laser geometry, recent work has shown that more complicated laser pulses and geometries or high atomic number targets could be favourable to cascades \cite{Gonoskov14,Vranic17,Gonoskov17,Tamburini18}.  Again we would not expect a more complicated laser pulse geometry to qualitatively change the cascade regimes.  The largest difference would be expected for a single laser illuminating the target from one side.  By comparing the results from a recent paper considering this case \cite{DelSorbo18} we see that the regimes are broadly similar but that the cascade occurs at much lower intensity in the counter-propagating laser case considered here -- for single-sided illumination the target is accelerated to relativistic speeds by the laser's radiation pressure, reducing the intensity in its rest-frame \cite{Kostyukov16,DelSorbo18} (although Doppler boosting can increase the degree of collimation of the emitted gamma-ray photons \cite{Capdessus18} and perhaps also the produced pairs, which may be advantageous for some applications).  Comparison to this work suggests that if dense pair plasma production is the desired outcome of an experiment then two-sided illumination of a near-critical density plasma is the ideal choice.  If cascade suppression is required then the choice should be single sided illumination of a significantly under-dense or over-dense target and in the over-dense case ions will be accelerated efficiently without laser energy loss to a cascade-produced pair plasma.

\section{Conclusions}

We have simulated QED cascades in the case of two counter-propagating circularly-polarised lasers of intensity $I\in [0.1,5]$\ Wcm$^{-2}$ interacting with a hydrogen plasma foil of thickness $1 \, \rm \mu m$ and initial density $n_0\in[10^{26},10^{31}]$\ m$^{-3}$. We found that above a threshold intensity the cascade saturates producing a pair plasma with density equal to or greater than the relativistically-corrected critical density.  The optimum target density for this was found to be the relativistic critical density at which substantial absorption  of the laser by the created pair plasma occurs ($\gtrsim50\%$).  For densities lower than this there are too few electrons in the in target to initiate a cascade (the number of pairs grows exponentially but does not reach saturation), at higher density the skin effect screens the laser fields and the cascade is suppressed. This provides a guide for pair plasma production experiments with next generation multi-PW lasers.

 \section*{Acknowledgments}

This work was funded by the UK Engineering and Physical Sciences Research Council (EP/M018156/1).  Computing resources have been provided by the York Advanced Research Computing Cluster.  The data required to reproduce the results presented here is available at doi:10.15124/5c9be63a-25e0-4e8b-b8ee-2781cc40b6fa.

\bibliography{article_NJP_CPR_final}

\end{document}